# A Secure Architecture for Standard Medical Imaging Repositories

Rui Lebre

IEETA
University of Aveiro
Campus Universitário de Santiago,
3810-193 Aveiro
Aveiro, Portugal

Luís Bastião

BMD Software
PCI - Creative Science Park Via do
Conhecimento, 3830-352 Ílhavo
Aveiro, Portugal

bastiao@bmd-software.com

Carlos Costa

IEETA
University of Aveiro
Campus Universitário de Santiago,
3810-193 Aveiro
Aveiro, Portugal

carlos.costa@ua.pt

University of A Coruña
Campus de Elviña, 15071 A Coruña
La Coruña, Spain

ruilebre@ua.pt

**Corresponding author:** Rui Lebre (ruilebre@ua.pt)

*Abstract -*

Background and Objective: Nowadays usage paradigms of medical imaging resources are requesting vendor-neutral archives, accessible through standard interfaces, with multi-repository support. Regional repositories shared by distinct institutions, tele-radiology as a service at Cloud, teaching and research archives, are illustrative examples of this new reality. However, traditional production environments have a server archive instance per functional domain where every registered client application has access to all studies. This paper proposes an innovator ownership concept and access control mechanisms that provide a multi-repository environment and integrates well with standard protocols.

Methods: A secure accounting mechanism for medical imaging repositories were designed and instantiated as an extension of a well-known open-source archive. A new Web services layer was implemented to provide a vendor-neutral solution complaint with modern DICOM-Web protocols for storage, search and retrieve of medical imaging data.

Results: The concept validation was done through the integration of proposed architecture in an open-source solution. A quantitative assessment was performed for evaluating the impact of the mechanism in the usual DICOM Web operations.

Conclusions: This article proposes a secure accounting architecture able to easily convert a standard medical imaging archive server in a multi-repository solution. The proposal validation was done through a set of tests that demonstrated its robustness and usage feasibility in a production environment. The proposed system offers new services, fundamental in a new era of Cloud-based operations, with acceptable temporal costs.

# A Cloud-ready Architecture for Shared Medical Imaging Repository

*Abstract*

Background and Objective: Nowadays usage paradigms of medical imaging resources are requesting vendor-neutral archives, accessible through standard interfaces, with multi-repository support. Regional repositories shared by distinct institutions, tele-radiology as a service at Cloud, teaching and research archives, are illustrative examples of this new reality. However, traditional production environments have a server archive instance per functional domain where every registered client application has access to all studies. This paper proposes an innovator ownership concept and access control mechanisms that provide a multi-repository environment and integrates well with standard protocols.

Methods: A secure accounting mechanism for medical imaging repositories were designed and instantiated as an extension of a well-known open-source archive. A new Web services layer was implemented to provide a vendor-neutral solution complaint with modern DICOM-Web protocols for storage, search and retrieve of medical imaging data.

Results: The concept validation was done through the integration of proposed architecture in an open-source solution. A quantitative assessment was performed for evaluating the impact of the mechanism in the usual DICOM Web operations.

Conclusions: This article proposes a secure accounting architecture able to easily convert a standard medical imaging archive server in a multi-repository solution. The proposal validation was done through a set of tests that demonstrated its robustness and usage feasibility in a production environment. The proposed system offers new services, fundamental in a new era of Cloud-based operations, with acceptable temporal costs.

## 1. Introduction

In the last years, digital medical imaging has seen its presence in healthcare providers ascending, following the evolutionary tendencies in the IT sector. Healthcare institutions have been constantly increasing the number of services provided for the patient's well-being like, for instance, telemedicine or integrated electronic Patient Record (ePR) [1]. Nowadays, remote reading of medical imaging studies is likely to be one of the most successful health services [2]. It is characterized by the transmission of medical images (CT, XA or CR, for instance) over the network and its further interpretation for diagnostic purposes.

The communication processes, data format, storage, querying, retrieval, visualization and printing of such medical imaging information are specified by the international Digital Imaging and Communications in Medicine (DICOM) standard [3], the leading normalization effort in this area [4, 5, 6]. The information about the patient is agglutinated in one or multiple standard files containing, besides image pixel data, metadata related to the patient, study, or report.

Picture Archiving and Communication Systems (PACS) are systems responsible for the acquisition, storage and distribution of the medical images [7], alongside with the managing of the workflow, that deeply rely on DICOM standard. A typical infrastructure is composed of one or multiple archives, acquisition modalities (data production units), distribution mechanisms and visualization equipment (workstations). In a traditional production environment, the archive serves a single organizational domain where the authorized users have access to all resources within the repository [8]. In these scenarios, accounting mechanisms like Role-Based Access Control (RBAC) mechanisms are not required. Nevertheless, this reality is also changing since the biggest organizations are being divided into departments with different areas of activity. Besides, there is also the need for division of infrastructures, i.e. multiple realms of data belonging to the same organization. Proprietary PACS solutions provide, since the beginning, some type of authorization mechanism that associates resources to user profiles. However, if a third-party client tries a DICOM Query and Retrieval request in the archive, those rules are not applied. In a traditional PACS, an external client (with access to C-FIND and C-MOVE services) can query and retrieve all DICOM object stored in the archive.

The exclusivity of inter-institutional processes is also changing with the creation of regional archives [9, 10] or even outsourcing of storage services that runs at Cloud [11, 12, 13] and serves many users and domains. Healthcare institutions are continuously investing in better IT infrastructures to support medical imaging departments, including installation and maintenance. Traditional operational business model and methods are becoming obsolete with the proliferation of reliable cloud storage services and its adoption by healthcare in detriment of local infrastructure [14]. The new technological advances are opportunities to develop new products and services that change the way how data is communicated and shared, anytime and anywhere, at high speed. The main advantages of cloud computing are cost savings, wide availability and high scalability [15, 16, 17, 18]. Nowadays, we are at an early stage transition of the medical imaging market to the Cloud.

This article proposes an architecture to support multi-user and multi-archive paradigm on a standard medical imaging repository, without interfere in the regular workflows supported by DICOM. The main goal is to provide resources ownership mechanisms and protection from unauthorized access over HTTP or HTTPS by providing an access control services to external entities. Those resources may be image objects, management or organizational services. A security layer to act directly in DICOM Web services (i.e. WADO-RS, QIDO-RS and STOW-RS) was developed, alongside with a layer and a web interface to manage the system users, resources and permissions in which the user credentials must be provided and verified before any operation conclusion. The proposed mechanism was integrated with a popular open-source archive, i.e. Dicoogle, validating the solution and demonstrating the advantages of having a multi-archive with a policy of multi-permission granting in the context of open medical imaging environments.

## 2. Background
### 2.1. Medical Imaging Systems
DICOM is the de facto standard in the medical imaging area [6, 19]. It is universally used to support systems interoperability, defining a non-proprietary medical data interchanging protocol, data format, file structure for medical images and its associated metadata, client and server services, workflow, among others [20].

The standard tries to model the real-world scenario in healthcare institutions. This is defined in the DICOM Information Model (DIM) [21] that deals with the structure and organization of the information related to the medical image. In this model, a patient can have one or more studies. Each study can have multiple series. A series assimilates a modality (or type of data) and contains one or various DICOM object instances. DICOM objects contain a wide range of metadata related not only to the image itself but also about the acquisition process, patient or specimen, or even organization.

Concerning the standard network layer, DICOM defines a set of services (DIMSE services), designated to handle the storage (C-STORE), query (C-FIND) and retrieve (C-MOVE) of DICOM objects over TCP/IP. Later, with the rising popularity of the Web Service technologies, the DICOM Web version was released, making the enumerated DIMSE services available through the web (i.e. accessible by HTTP). WADO-RS (Web Access to DICOM Objects by RESTful Services) is responsible to make available the access to DICOM objects like images or reports. It enables the retrieval of a specific study, series, instance or frames through HTTP. QIDO-RS (Query based on ID for DICOM Objects by RESTful Services) enables the searching for studies, series or instances and their requested attributes. Finally, the STOW-RS (Store Over the Web by RESTful Services) allows the storage of DICOM objects in the Web archive server. The proliferation of DICOM compliant equipment enabled the exchange of data between medical imaging devices and triggered the implementation of PACS [22]. It contemplates a set of hardware and software that processes, stores, distributes and provides medical images or a portion of them in the healthcare institution [3, 23]. For instance, modalities, digital image archives, workstations or printers are components belonging to PACS.

Integrating the Healthcare Enterprise (IHE) is an initiative that promotes a standard-based approach for developing universal-accepted solutions and, thus, providing interoperability [24, 25].

XDS (Cross-Enterprise Document Sharing) enables the sharing of patient information. This is a document-centric architecture that allows query and retrieve specific data for a specific patient [26]. XDS model incorporates and central registry that store metadata relative to the published document. IHE introduced XDS-I.b (Cross-Enterprise Document Sharing for Imaging) [24]. This profile addresses a solution for sharing documents across multiple institutions [26]. The payload of those documents may encompass imaging studies, including acquired images, diagnostic reports or a selection of relevant images related with the report requested [26, 27].

*2.2. Dicoogle*

Dicoogle is a popular open-source PACS archive [28] that has a modular architecture and uses a document-based indexing system as database [23]. Its plugin concept and the provision of a software development kit (SDK), encourage developers and researchers to quickly develop a new functionality (see Figure 1). Dicoogle can be used to support three distinct usage scenarios, taking advantage of the modular architecture: production, research and teaching.

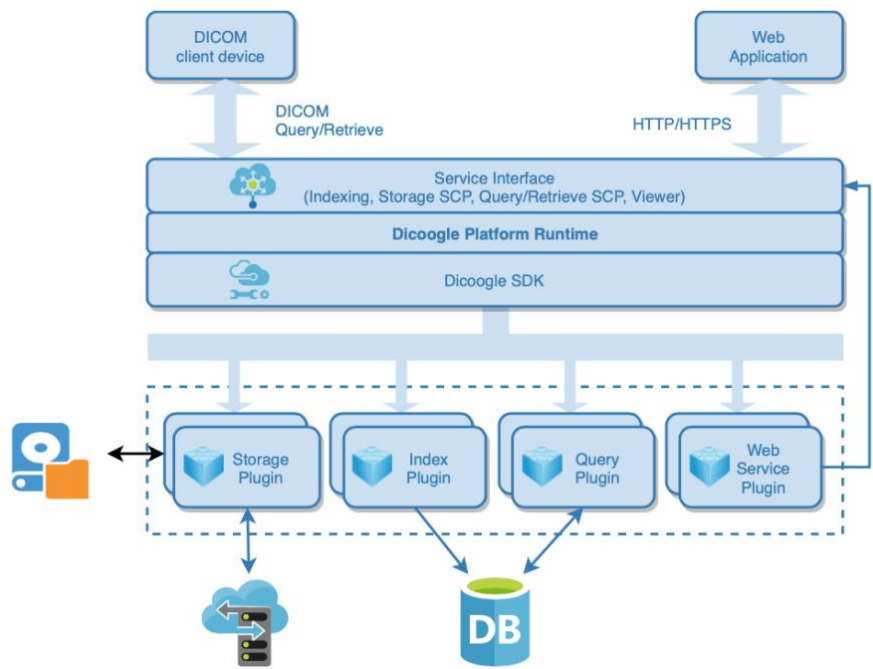

*Figure 1 - Dicoogle general architecture. Adapted from [29].*

A Vendor Neutral Archive (VNA) is a medical imaging technology in which DICOM objects (e.g. images) are stored in a standard format with a standard interface, and they can be accessed in a vendor-neutral manner by third party systems. Dicoogle fits this definition since the archive stores the objects in DICOM format and provides DICOM network interface for storage, query and retrieval. Its VNA characteristics allow to connect external components as visualization software, modality worklist or other complementary services.  In the context of Cloud-based operations, the Dicoogle services are consumed using the DICOMWeb interface, i.e. QIDO for content discovery, WADO for content retrieval, and STOW for storage. The proposed RBAC mechanism works over this Web interface for supporting multiple repositories in a single server instance of Dicoogle.

Dicoogle SDK was created in order to simplify the development of new features [28] by third parties while still assuring compatibility with the core functionalities. After this process, Dicoogle will automatically load the new modules on startup. Dicoogle SDK makes immediately available all operations related to storage, querying and indexation via its internal API [29]. Plugins may be of 5 categories: storage, index, query, web service and web user interface. Storage plugins are responsible for handling the store and retrieval of DICOM data. The implementation of this type of plugin would either keep the files in a local filesystem or in a remote storage. Index plugins handle the indexation of the metadata extracted from the DICOM objects archived. The indexation of the files allows the quick search and retrieval and it is closely interconnected with another type of plugins, the query plugins. As the name says, query plugins allow to query the indexed data. Usually, the query provider is coupled with a particular indexer and even bundled in the same plugin set. It is also possible to extend the web services present in Dicoogle, developing the Web Service plugin. Besides the extension of new services, developers may also implement new UI (User Interface) components that are automatically loaded into Dicoogle's web application, the Web User Interface plugins.

Besides SDK, Dicoogle offers other resources for supporting the development of new features. Dicoogle Learning Pack[1] and more recently, the Dicoogle API[2] definition, has published a set of learning resources as guides, definition of the technical API and definition of the Dicoogle default web services. These guides assist the learning process of good practices while developing plugins, as communication between components or logging of information.

*2.3. Access Control in Healthcare Information Systems*
The literature reports some works regarding privacy and security in medical repositories and healthcare environments. In fact, many authors [30, 31, 32, 33] find the usage of access control and essential tool in healthcare. Fabian et al. [33] assert that inter-organizational sharing and collaborative use of medical data has becoming very important. The authors propose in the same paper an architecture and implementation of a mechanism to securely share healthcare patient data.

In [2], the author presents and discusses principles and requirements already developed for electronic health records, focusing after on privacy protection requirements, controls and security services including accountability. Among other requirements, it highlights that systems must certify that data is not disclosed or accessed by unauthorized users during the session(s). It is recommended that employees inside the healthcare institution should be under access control services, ensuring that no unauthorized agent have access to data.

In [34], the authors empathize the importance of the modernization of the healthcare system. However, with the development in this area, some challenges come across regarding security, safety and privacy of patient's records. The authors question how to provide a secure way of sharing medical data. The solution is assured by access control, which verifies the person's access permission in order to ensure security.

Ma, W. et al. [8] focus on the nowadays issue that everyone that logon the workstation in the lab can access every study of every patient. The author also warns that once the client application entity (IP, port and AETitle) are added to the PACS whitelist, any user logged on as a client application has access to all images stored in the archive. Later on, the same authors [35] propose a security middleware infrastructure. The main objective was to develop an infrastructure that provides online security mechanism such as authentication and authorization, audit logs and dynamic setting of access rights by authorizing user's behavior.

According to [32], most recent access control systems are very inflexible when using role-based access control schemes. In [36], Seol, K. et al propose an attribute-based access control model that can be flexible when comparing to the existing RBAC schemes.

Mashima and Ahamad [37] proposed an architecture to control the access of the patients to the health information. Their mechanisms focus on when and how the information can be accessed. In [38], authors combined RBAC with Flexible Authorization Framework (FAF) to specify access control policies, a framework where authorizations are specified in terms of a hierarchic rule-based logic.

More recently, in [31], the authors propose Cerberus, an access control scheme to enforce access to Cohort Study Platforms (CSP). It ensures the full stack of the access right management using ontologies and flexible access control.

---

[1] URL: https://bioinformatics-ua.github.io/dicoogle-learning-pack/

[2] URL: https://bioinformatics-ua.github.io/dicoogle-api/

In [39], the authors stand that security is an essential issue in telemedicine. However, in spite of the existing solutions from nowadays, there is a lack of an access control mechanism prepared to deal with the DICOM information model and objects integrated in a PACS environment.

## 3. Methods

### *3.1. Access Control Model*

The proposed system implements a role-based access control (RBAC) model. Its definition started with the analysis of production PACS environment in order to characterize the entities, resources, processes and relationships. Figure 2 illustrates the data model of the solution, representing the production environment. In the real world, a User belongs to a Facility in an Organization. Subsequently, the Organization/Facility produces, stores or distributes Resources. Exemplifying, those Resources may be medical equipment, DICOM objects or reports. Moreover, the User shall have the possibility to access to each Resource according with defined Permissions. To simplify the attribution of Permission, it was decided to create the entity Role that is responsible for aggregating Permissions of multiple Users like, for instance, Administrative staff, Cardiologists, Neurologists, etc. The creation of previously described entities will allow the development of a platform with access control functionalities. The authentication and authorization will only be given by default to authorized personnel belonging to the same organization which owns the requested resource. Additionally, authorization will only be given to personnel who have the permission to perform the exact action (i.e. READ, WRITE, CREATE). Furthermore, it is possible to share resources or permissions. A User that has access to a Resource and granted the SHARE action permission, may share that Resource with other users.

The proposed model is an abstraction of a real-work medical imaging laboratory environment. However, it can be applied in other situations. Figure 3 is shown the example of the usage of the framework in an academic scenery. On top of the hierarchy is the Organization. In this case, the organization shall be the University. Bellow the hierarchic position of the organization, the Schools of Computer Science and Health represent two of the various Facilities the Organization can have. Belonging to the schools, there is the Students A, B and C, the Professor and the Resources 1, 2 and 3.

In the environment depicted in the Figure 3, the professor, belonging to the University and owning the full right of access to both the Schools, has the competency to give and revoke permissions to students. Focusing on Student A, he belongs to both the Facilities School of Computer Science and School of Health. If Student A is given permissions to READ resources, he will be able to access the resources. However, Student B, since she belongs only to the School of Computer Science, by default, she just has access to Resource 1 from the School of Computer Science. On the other hand, Student C can have access, by default, to Resource 2 and 3. All of the permissions are managed by the professor, the actor responsible for setting the default definitions of the archive. I.e., the Professor, in this environment, is responsible for setting if the students have permission to all the resources of their environment or, contrarily, revoking the access to, for instance, DICOM Objects of a specific Modality.

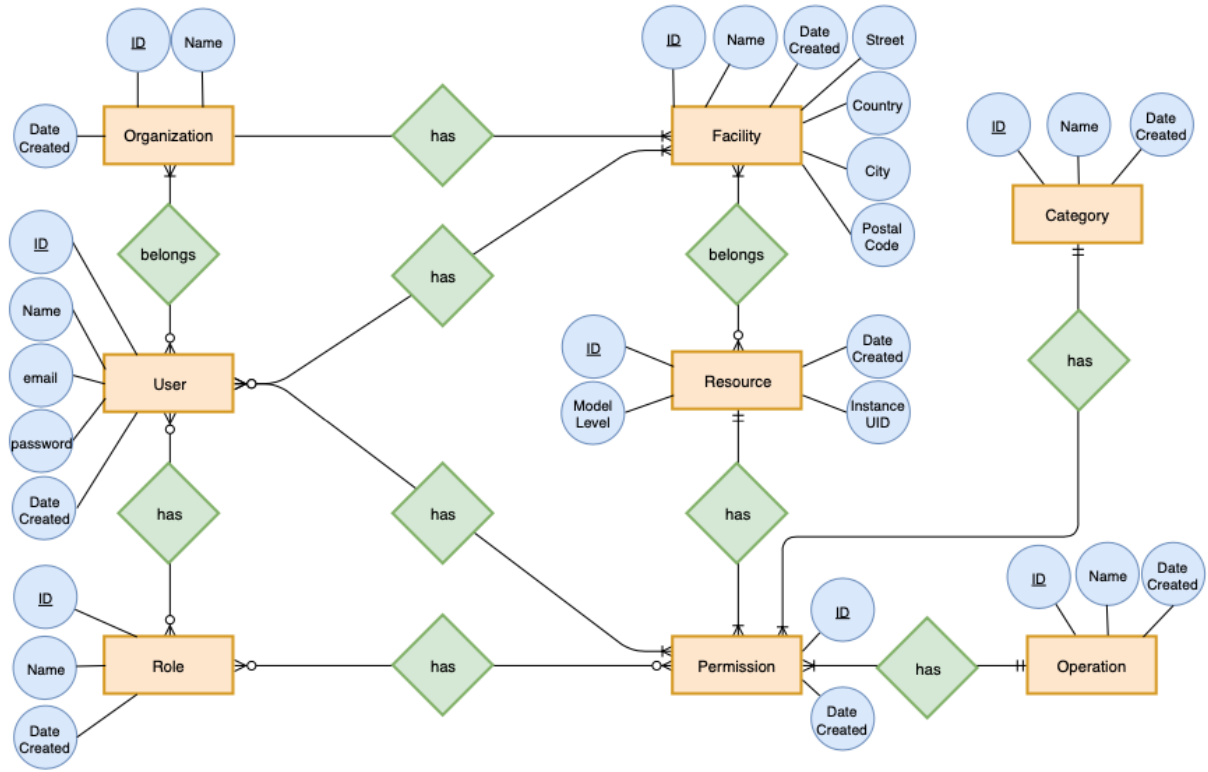

Figure 2: Representation of the *Role Based Access-Control* UML data model.

Furthermore, the framework open doors to restrict the access to resources to some slots in the timetable. In the schema of the Figure 3, for instance, the access permissions to the resources by the Students A, B, and C could be restrained to only the time of the class, revoking the access at the moment the class finishes.

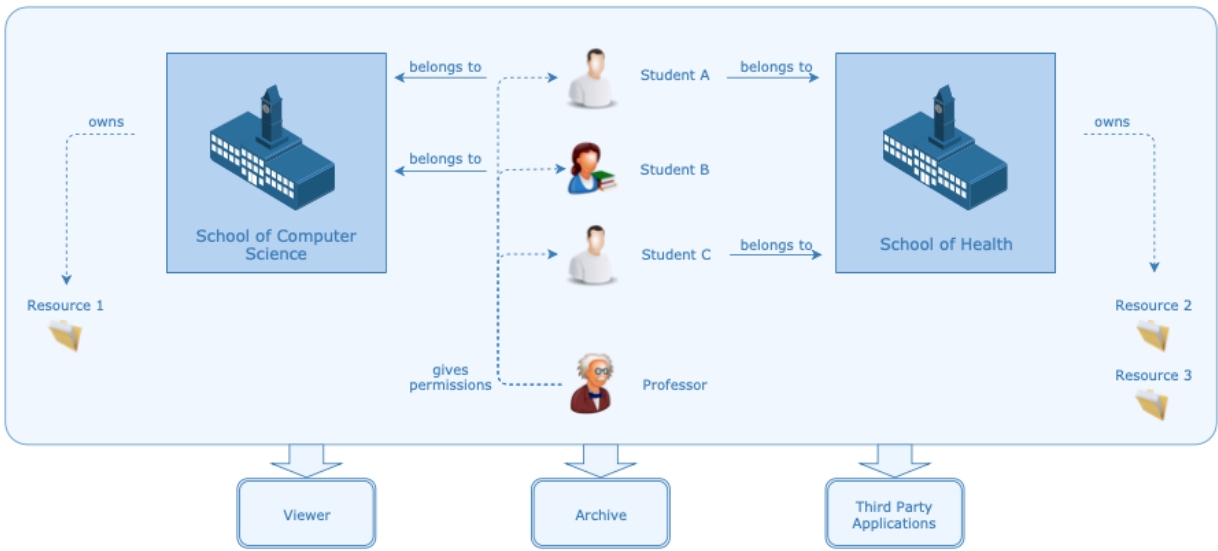

Figure 3: Representation of the *Role Based Access-Control* model in the academic environment.

*3.2. Architecture Overview*

After defining the access control model, it was necessary to implement it in a DICOM standard structured repository for restricting resources access to authorized users, providing mechanisms to manage entities, resources and roles. As expressed, our proposal was instantiated as an extension of Dicoogle Open-Source PACS that already provides several features to maintain a standard archive and for extraction of metadata associated with every DICOM object. The Dicoogle SDK was used to develop and integrate the RBAC module with Dicoogle core functionalities (Figure 4). Moreover, other Dicoogle modules can also consume the new security services provided by the RBAC module. The development of the described accounting mechanism was particularly motivated by the necessity of offering those security services to the DICOM Web interface. In this context, Dicoogle WADO-RS, STOW-RS and QIDO-RS services were re-factored to consume the new RBAC module. The activation of this module was defined as optional and managed by Dicoogle administrators. In a Dicoogle instance with this RBAC option active, a third platform will consume the Web Dicoogle services (storage, query and retrieve) with the access to resources filtered by the RBAC policies. In this case, the RBAC User is the one provided in the HTTP authentication layer. Finally, a web portal was also developed to support end-user management of proposed architecture. It consumes a new set of services available through a Rest API, developed as well as a plugin. This new interface allows managing Organizations, Facilities, Users, Permissions, Resources, permission sharing and login/logout. The next sections will describe in detail the main developed modules and its integration with Dicoogle services.

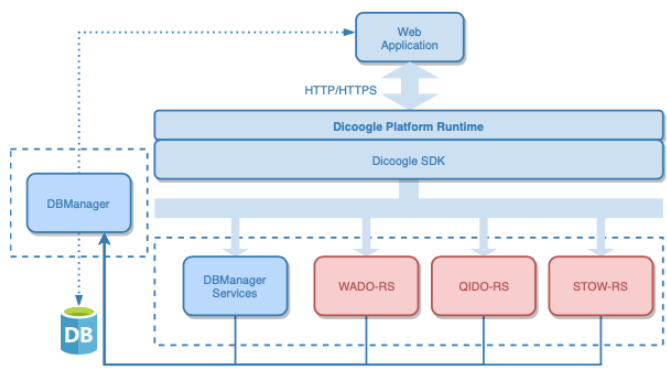

Figure 4: Proposed architecture - Dicoogle Integration

*3.3. DBManager and Associated Services*

Dicoogle open-source PACS already supports different users, some with administrator privileges. However, the system does not allow a different specification per user, like a READ or WRITE permissions or restricted access to some modalities. Moreover, the users are not integrated with Dicoogle DICOM services for authentication and authorization purposes. To support the requirements of the new RBAC module, namely permissions granting or denial for reading or writing objects, it was created an intermediate module to be included in the Dicoogle's architecture. As shown in Figure 4, a new DBManager module was appended to the Dicoogle architecture and it is responsible for modeling the system entities along with DICOM data model: Patient, Study, Series and Image. DBManager module will be in strict communication with an external Database Management System (DBMS) responsible to store and index all the permission granting to every object or device within healthcare domain.

DBManager can be also integrated with healthcare information system policies, for instance, through LDAP or HL7 protocols. LDAP makes possible to centralize the user/role management in a single place, for the convenience of IT departments. For instance, it may be configured to fetch the user's profiles and configure the mapping to a specific authorization role. For the patient and its data management, the HL7 manager module plays a significant role because it allows to define specific trigger actions to update the patient data when they are inserted into database, keeping this way the consistent of data across different systems. Moreover, if the patient data is updated, the system allows to receive message for update/merge information.

Finally, DICOM web modules communicate continuously with DBManager for checking the identity and permissions of every service requester.

In order to provide an interface for external developers, it was developed a set of services for interaction with the DBMS through DBManager, made available through a REST API. All the services were developed as a Dicoogle plugin (Jetty service plugin) and the main goal was to have access to the basic CRUD functions of the proposed access control mechanism system. Those services must be protected against unauthorized accesses. To do so, a filtering strategy was used (Figure 5), a protecting barrier between the operation requester and the operation itself. Once the reception of the request takes place, the filter checks which permissions are needed to perform the requested operation.

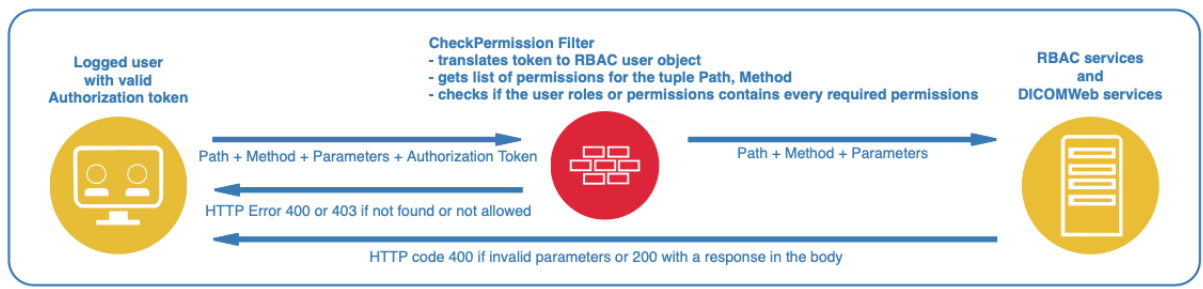

Figure 5: Filter usage scheme

The requests have an HTTP header with attribute "Authorization" that will contain the Dicoogle session token, represented by an alphanumeric string of characters. A connection to Dicoogle's core is made to get data about the token such as the user that is logged in. Since there are distinct accounting systems, it is required to verify, first of all, if the user is a member of the access control system. Next, the system will check if the user's roles grant permission to perform the requested operation, including the case where the permission is shared by another user. After executing the authentication and authorization process, three results are possible: 1) user is authorized to perform the request and the normal service flow is started; 2) user is not authorization to perform the operation and a 403 (Forbidden) HTTP error message is returned; 3) user is accessing to an invalid service and a 400 (Bad Request) HTTP error message is returned.

*3.4. DICOM Web with RBAC*
Some changes were introduced to the Dicoogle implementation of DICOMWeb standard to support the developed RBAC mechanism. As described in the last section, the HTTP service request must contain the "Authorization" attribute in the header to make use of proposed system. If not provided, the system may only provide access to public resources or simply deny the service. Therefore, using the DBManager module, the system will perform the validation of the authorization. Figure 6 illustrates the sequence diagram of WADO-RS service, since the Dicoogle user login until the DICOM object retrieval. On step 1, the user provides the username and password to Dicoogle platform, that returns the access token if the authentication is valid (step 2). Once the client application has the token, the WADO-RS request (step 3) includes this token in the HTTP header. This token is used by WADO-RS service to retrieve the user information from Dicoogle core platform (steps 4 and 5). Then, WADO-RS service establishes communication with the DBManager module in order to check the user authorization to resource access (steps 6 and 7). If the permission is granted, the subsequent steps are the same as in a regular WADO-RS service operation (steps 8 to 12). The grant is only provided if the user has GET operation permission on the requested object RESOURCE.

STOW-RS service logic is very similar to WADO-RS, where the user also has to obtain a session token an make the request with that session token in the Authorization header. However, the permissions are slightly different, being necessary to have operation permission "ADD" of a RESOURCE category.

Finally, QIDO-RS service implementation requested some additional changes relative to the processes implemented in the WADO-RS and STOW-RS cases (Figure 7). Analogously to the previous cases, the third-party client application shall perform the login in the Dicoogle PACS and get the session token to use the service. However, QIDO-RS plugin does not check the permission rights of the user. The associated token is broadcasted to all the index/query plugins of the Dicoogle (in the current example of Figure 7, step 4). The task to check the rights is performed by the index/query plugins that support the accounting mechanism framework (step 5 to 7), once every single index/query plugin needs to check if the user has permissions. The query is then performed, and the results truncated to include only the records that the requester has permissions to LIST operation in RESOURCE category.

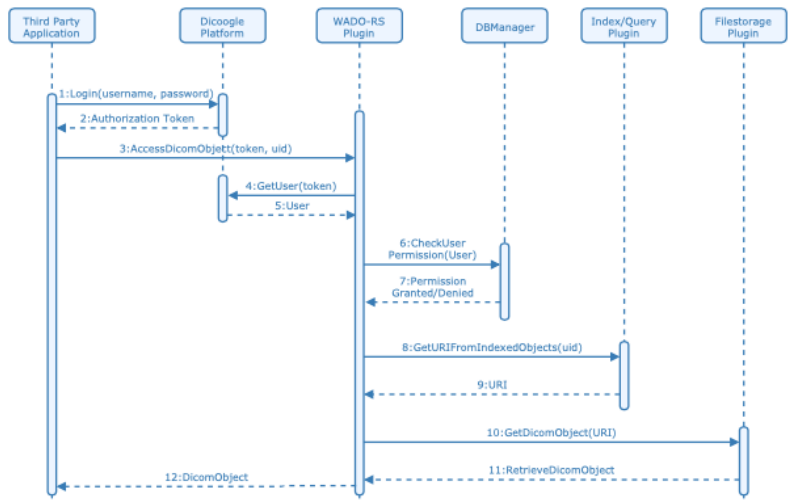

Figure 6: Filter usage scheme for WADO-RS DICOM Web service

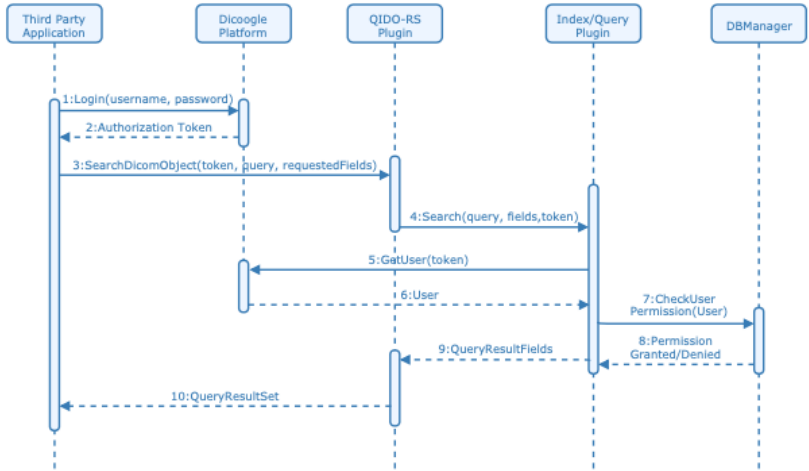

Figure 7: Filter usage scheme for QIDO-RS DICOMWeb service

*3.5. Logging and Security Auditing*

Research shows that access control is not a complete solution for securing a database [40] and must be incorporated with other mechanisms that provide complementary services since most of the breaches are done by insiders. Logging and auditing are important complementary mechanism. Logging is the record of all users request and activities. Auditing is the process of review (analysis) this data to ensure that authorized user do not misuse their privileges.

Dicoogle Platform Runtime, which component is represented in Figure 1, logs all the operations related to its plugins. The operations in the developed modules follow the directives present in the Dicoogle Learning Pack. The logging of the operations allows further audit and traceability of the requests and operations. The audit trail mechanism registers most of the operations for each user, that are associated with the organization and facility, including the resources consumed and accesses denied. It is recorded the timestamp, request URL, parameters, category, operation, user agent and also IP address. The data is collected in a separated database and the review process is called security auditing. The data itself comprises the audit trail. The data can be reviewed by authorized users, using a developed web application (Figure 8), to verify if archive is being used in accordance with the healthcare provider's data security requirements.

Figure 8: Web application for archive security auditing

## 4. Validation

The user accounting mechanism introduces a delay on the regular services execution, i.e. when compared with Dicoogle without RBAC layer active. The impact is explained by the addition of additional processes like the need to verify, for each HTTP request, the authorization clearance of the entity performing the request. Concerning the validation of proposed system, what is pertinent it to assess if the delay is not relevant (i.e. acceptable) in the context of production environments. Therefore, several tests were made to evaluate the performance impact of the proposed solution. The methodology used simulates the massive usage of Dicoogle DICOM Web services (STOW-RS, QIDO-RS and WADO-RS), with and without RBAC module active.

*4.1. Assessment Procedure*

The validation tests were carried out in a virtual machine running Linux (Ubuntu 16.04.5 LTS). The virtual machine specifications encompassed an Intel X (R) CPU E5-2630 v4 @ 2.20GHz CPU with 4 cores available, memory of 4GB and 55GB of ROM memory.

In order to ensure maximum testing equality, some aspects have been considered, such as:
• Same software version including operating system and DBMS (MySQL 5.7.17);
• Same amount of memory;
• Same DICOM objects dataset;
• Same system load and number of active applications and services;
• Similar test replications, being the authorization and provision of the access token, the only factor changed.

The tests aim to validate the DICOM Web services (QIDO-RS, WADO-RS and STOW-RS) robustness and performance when using the RBAC module. So, tests were carried out in the Dicoogle with and without authentication. The procedures are identical, having only one change that is the addition of the session token to each request with authentication active. Before the service tests start, a request was made to the login service in order to obtain the service access token. It was necessary to create a test user with appropriated permissions, namely full permissions to store DICOM objects (to test STOW-RS), randomly generated permissions to query DICOM objects (to test QIDO-RS) and randomly generated permissions to access DICOM objects (to test WADO-RS). This user was also associated to one organization and 5 facilities.

In the case of QIDO-RS service, queries were built to try each level of the DICOM Information Model: Patient, Study, Series and Instance. The search parameters included several attributes such as Modality, SOPInstanceUID, SeriesInstanceUID, StudyInstanceUID and PatientID. The tests were repeated 500 times.

Regarding the STOW-RS service, the storage of 5675 DICOM Objects was performed. The tests were repeated 5 times. The distribution of the number of files per size is described in Table 1.

| Number of files | 2224 | 1120 | 960 | 417 | 368 | 352 | 156 | 78 |
|---|---|---|---|---|---|---|---|---|
| Size (Kilobytes) | 131 | 290 | 163 | 132 | 514 | 394 | 515 | 130 |

Table 1: File size of each DICOM file

To evaluate the WADO-RS case, 900 identifiers were used for objects retrieve purposes. Using those identifiers, sequential tests started to be performed, requesting the first frame of each DICOM object in which all the files were in the permission list of the testing user. The tests were repeated 4 times.

For the entire set of tests, the Unirest library was used since it allows the invocation of REST services. The temporal duration of service, from the request creation until the request answer, was recorded. Values such as Status code of the response, name of the file to be stored (STOW-RS), query (QIDO-RS), SOPInstanceUID (WADO-RS), wait time and response body were saved in a log file.

*4.2. Evaluation Results*

The results of the performed tests are shown in Table 2. The table shows a comparison between the Dicoogle PACS DICOMWeb plugins with open access and with RBAC controlled access. As expected, there was a growth in the services time of execution: 4,25 times for STOW-RS; 3,10 times for QIDO-RS; and 2,45 times for WADO-RS.

|  | STOW-RS | QIDO-RS | WADO-RS |
|---|---|---|---|
| Open access (ms) | 36 | 21 | 42 |
| Protected access (ms) | 153 | 65 | 103 |

Table 2: Impact of the mechanism in the usual DICOM Web operations

As explained before, the execution time increased due to the additional logic and database operations to check the user permissions for a specific resource. Although the overhead of 117 ms, 44 ms and 61 ms in the cases of STOW-RS, QIDO-RS and WADO-RS, respectively, the total running time presented in the table is still very acceptable in a production environment. The individual overhead in each case does not conduct to major loss if there are more batch operations to run in the database, such as the User belonging to 100 Facilities, for instance.

Particularly considering the STOW-RS plugin, each time a request is made to store a DICOM object, it is necessary to create the new resource in the database and define associated permissions. The permissions were created to the user that stored the DICOM object but also to all the facilities he belongs. In the test case, the user belongs to 5 facilities.

In the QIDO-RS case, the SQL query of the index/query plugin had to contain additional restrictions to check the eligibility of the user to retrieve the results. This means that the results retrieved from the Database Management System are already filtered.

Finally, the WADO-RS has the lowest delay. The reason for this is explained for the lower number of permission checking and database requests number.

## 5. Conclusion

VNAs are fundamental in modern medical imaging environments since they use standard data format and interface [41, 42], making images accessible to healthcare professionals regardless of what proprietary system created them. Traditional PACS solutions usually provide some type of proprietary authorization mechanism in their visualization workstations that associates resources to user profiles. However, if a third-party client application has access to the archive through DICOM standard network interfaces, those rules are not applied. As result, the external user can query and retrieve all DICOM object stored in the archive without control.

Other constraint of traditional PACS archive architectures is that they only support a functional domain repository per single server instance. However, nowadays Cloud-based storage paradigm is requesting multi-repository archive server or, in other words, a service accounting for DICOM standard archives. For instance, regional repositories shared by distinct institutions, tele-radiology as a service running in a Cloud provider, teaching and research archives, are illustrative examples of this new reality.

This article proposed a secure accounting mechanism for medical imaging repositories that can be integrated with modern DICOMWeb services for storage, search and retrieve medical imaging data. The solution was instantiated as an extension of Dicoogle open-source PACS. Besides providing accounting services for standard services, it also includes Web services API for integration with external applications and development of new services in the future. The proposal validation was done through a set of tests that demonstrated its robustness and usage feasibility in a production environment. The conclusion is that the proposed system offers new security services with acceptable temporal costs.